\documentclass[aps,twocolumn,a4paper]{revtex4}
\usepackage{epsfig}
\usepackage{graphicx}
\usepackage{amsmath,amssymb,color}
\usepackage[english]{babel}

\parskip=\medskipamount

\newcommand{\eq}[1]{(\ref{#1})}
\newcommand{\fig}[1]{Fig.~\ref{#1}}

\newcommand{\be}{\begin{equation}}
\newcommand{\ee}{\end{equation}}

\newcommand\disp{\displaystyle}

\newcommand{\im}{\textrm{Im}\,}

\begin{document}

\title{Buckling and wrinkling from geometric and energetic viewpoints}

\author{S. Nechaev$^{1,2}$ and K. Polovnikov$^{3}$}

\address{$^1$Universit\'e Paris-Sud/CNRS, LPTMS, UMR 8626, B\^at. 100, 91405 Orsay, France \\
$^2$ J.-V. Poncelet Laboratory, CNRS, UMI 2615, 11 Bolshoy Vlasievski, 119002 Moscow, Russia \\
$^3$Physics Department, Moscow State University, 119992, Moscow, Russia}

\date{\today}

\begin{abstract}

We discuss shape profiles emerging in inhomogeneous growth of squeezed tissues. Two approaches are
used simultaneously: i) conformal embedding of two-dimensional domain with hyperbolic metrics into
the plane, and ii) a pure energetic consideration based on the minimization of the total energy
functional. In the latter case the non-uniformly pre-stressed plate, which models the inhomogeneous
two-dimensional growth, is analyzed in linear regime under small stochastic perturbations. It is
explicitly demonstrated that both approaches give consistent results for buckling profiles and
reveal self-similar behavior. We speculate that fractal-like organization of growing squeezed
structure has a far-reaching impact on understanding cell proliferation in various biological
tissues.

\end{abstract}

\maketitle

\section{Introduction}

Variety of shapes of growing objects emerges often due to incompatibility of internal growth
protocol with geometrical restrictions imposed by embedding of growing substances into the space.
For example, buckling of a salad leaf can be naively explained as a conflict between growth due to
the reproduction of periphery cells, and geometrical growth of a circumference of a planar disc. It
is known that there exists a biological mechanism which inhibits a cell growth if the cell
experiences sufficient external pressure. So, cells located inside a body of growing tissue,
preserve their size, while periphery cells have less steric restrictions an proliferate easier.

If the circumference, $P$, of a two-dimensional membrane (tissue) of typical radius $R$ grows
faster than $P(R) \sim R$, one could expect buckling since "extra material" goes out of plane and
finds space in third dimension. The same is valid for growing 3D objects: suppose that growing
surface, $S$, of a tissue (for example, a spherical cell colony of typical size $R$) grows faster
than $S(R) \sim R^2$. In this case however, the expected colony shape is more cumbersome than for
membranes, because the "extra material" cannot go out in a forth dimension: our world is
three-dimensional and the grows is accompanied by surface instabilities.

Certainly such naive geometric viewpoint does not take into account bending energy of deformed
elastic tissues and definitely should be modified by an appropriate account of stresses existing in
non-uniformly deformed elastic two-- and three-dimensional objects. In this work, we restrict our
consideration to growth of 2D objects only. Currently, two groups of works, devoted to the
determination of shapes of growing bodies, can be distinguished in the literature.

In the first group of works, tentatively denoted as "geometric" \cite{1,2} (see also \cite{3}), the
determination of the typical profile of 2D surface is dictated by optimal "isometric" embedding of
discretized 2D surface into 3D space, and is described by an appropriate change of conformal
metrics. This approach relies only on geometric constraints imposed by a "conflict" between metric
of a 3D Euclidean space and intrinsic hyperbolic metric of a 2D ideal membrane, and can be treated
using the conformal methods. It should be mentioned, that formation of wrinkles within this
approach seems to be closely related to description of phyllotaxis via conformal methods
\cite{levitov}. The discussion of possible connection between wrinkling patterns and energy relief
of continuously deformed periodic 2D lattices of repulsive particles, will be considered
separately.

In the second, more numerous group of works \cite{hebrew,lewicka,gemmer,marder,ignobel,swinney},
conventionally named "energetic", more realistic approach, based on minimization of a bending free
energy of non-uniformly deformed thin elastic plates, is realized. Authors of almost all works,
devoted to buckling of 2D growing tissues, state that "geometric" and "energetic" approaches are
complimentary to each other. Besides, explicit comparison of results of these two approaches is
still absent, as well as there is lack of analytic results obtained within the "energetic
approach". In the current work we would like to fill this gap and consider from conformal
(geometric) and energetic viewpoints wrinkling and buckling, paying attention to similarity and
differences of obtained results.

The paper is organized as follows. In the Section II we discuss wrinkling in frameworks of a
conformal approach and explicitly define the surface profile under the constraint of elementary
surface area conservation. In Sections III-IV we consider buckling of non-uniformly squeezed thin
plate obtained by minimization of its bending energy in linear approximation and summarize obtained
results. Technical details of computations are presented in Appendices A-C.

\section{Conformal wrinkling}

Surface wrinkling of constrained exponentially growing tissue can be tentatively described by the
following naive geometric model. Consider initial flat elementary domain schematically depicted by
the triangle $A$ in upper row of the \fig{fig:01}a. Suppose that this domain represents, say, a
linearly ordered colony of proliferating cells. If all cells divide independently and
unconstrained, the colony doubles after some given time slot and expands its linear size.

To mimic the colony perimeter growth, suppose that 2nd generation of initial sample $A$ consists of
two aligned copies of $A$. Repeating such duplication recursively, we arrive at a sequence of
discrete time slots, $1,2,...k$, where in $k$th generation the perimeter of a colony, $P$,
expressed in number of copies of $A$, is just $P(k)=2^{k-1}$. In absence of external confinement,
two consequent duplications of initial sample $A$ are schematically shown in the \fig{fig:01}a.

Now, to impose confinement effect which preserves the lateral size of a system, suppose that each
generation $k$ of a colony, experiences lateral squeezing: as larger $P(k)$, as more external
pressure should be applied to return the length of a colony, $P(k)$, to its initial size, $P(k=1)$,
-- as it is shown in the \fig{fig:01}a. Under such a squeezing the elementary domains become
essentially deformed: as larger $k$, as more compressed the elementary domains $A$ in a
corresponding generation -- see the \fig{fig:01}b. Impose now an "incompressibility condition" and
demand the domains $A$ to preserve their surface areas in all generations. To satisfy this
condition, the squeezed elementary domains should buckle in third dimension orthogonal to the plane
of figures.

\subsection{The model}

\begin{widetext}

\begin{figure}[ht]
\epsfig{file=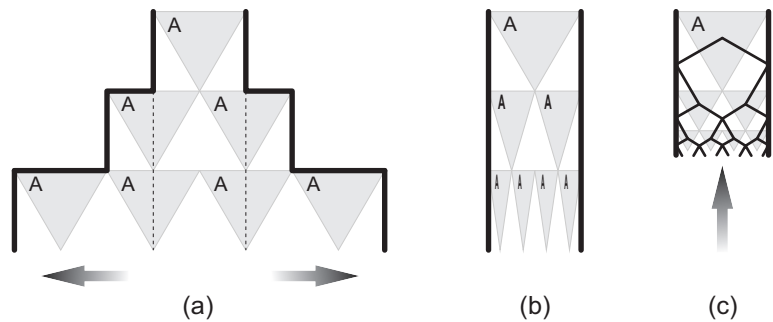,width=10cm}
\caption{Nonuniform squeezing of a sample consisting of duplicating elementary cells $A$:
a) Exponentially increasing lateral pressure is applied to each generation of samples $A$ to return
them to initial size shown by dotted line; b) Schematic illustration of result of nonuniform
(exponential) squeezing; c) Structure of a sample after nonuniform compression in vertical
direction. The sample is covered by the hyperbolic regular lattice.}
\label{fig:01}
\end{figure}

\end{widetext}

Let us reformulate the above hand-waving viewpoint on buckling in more rigorous terms allowing
analytic treatment. Make the nonuniform (exponential) discrete contraction of the sample shown in
the \fig{fig:01}c in vertical direction: the row $k$ is contracted by the factor $2^{k-1}$ along
the $y$-axis. Such contraction together with above defined squeezing gives the "discrete hyperbolic
tessellation" of a domain shown in the \fig{fig:01}c. The solid lines connecting the centers of
neighboring triangles constitute the hyperbolic graph embedded into this domain.

To make this hyperbolic tessellation consistent with the tessellation of Euclidean plane, we should
proceed as it has been prescribed in \cite{1}. Namely, take an equilateral triangle $A$ in the
complex plane $z=x+iy$ with Euclidean metrics, $ds^2=dx^2+dy^2$, and make a conformal mapping of
$A$ onto circular hyperbolic triangle $A'$ with angles $\{\alpha,\alpha,0\}$ ($0 \le \alpha
\le\pi/3$) lying in the upper half-plane $\im w>0$ of the plane $w=u+i v$ endowed with the
Hyperbolic metric $d\sigma^2=\frac{du^2+dv^2}{v^2}$. Let us denote this conformal mapping as
$z(w)$. Below we provide an explicit construction of $z(w)$.

\begin{widetext}

\begin{figure}[ht]
\epsfig{file=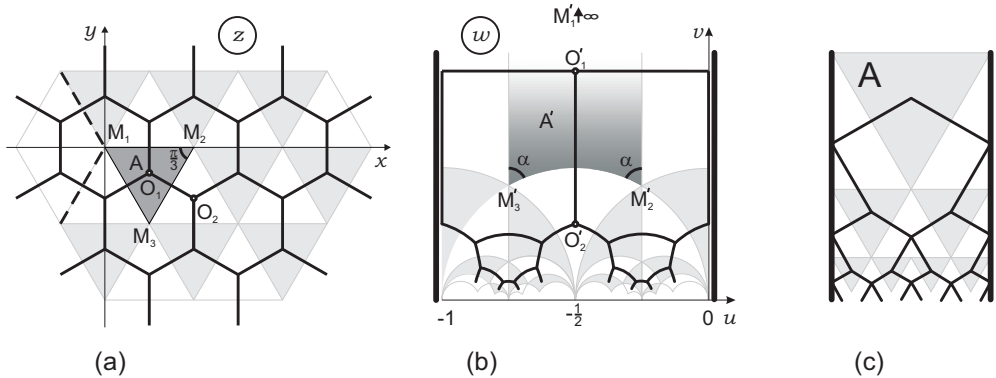,width=13cm}
\caption{Conformal mapping of Euclidean complex plane $z=x+iy$ to the half-strip $v>0$
($0\le u\le 1$) of the complex plane $w=u+iv$ with hyperbolic metrics. The mapping is defined by
the conformal transform of the equilateral triangle $M_1M_2N_3$ (a) to the circular triangle
$M_1'M_2'M_3'$ with angles $\{\alpha, \alpha,0\}$ (b). The figure (b) is drawn for specific value
$\alpha=\pi/3$. The figure (c) is the same as \fig{fig:01}c and and is drawn to emphasize the
similarity with discrete squeezing shown in the \fig{fig:01}.}
\label{fig:02}
\end{figure}

\end{widetext}

Other triangles in the Euclidean plain $z$ can be recursively obtained by reflections of the domain
$A$ with respect to its sides, then -- with respect to the sides of its images, \emph{etc}.
Correspondingly, we tessellate the Hyperbolic domain and cover \emph{isometrically} the strip in
the \fig{fig:02}b by the images of the equilateral triangle $A$. Remind that isometric covering
means that it preserves angles and distances in a given metrics.

It is easy to check that the conformal mapping $z(w)$ preserves the area of the hyperbolic
triangles. Namely, the area $S_A$ of the triangle $A$ in the the plane $z$ can be written as
\be
S_A=\int\limits_{\triangle A} dx dy
\label{eq:trian1}
\ee
where the integration is restricted by the boundary of the triangle. After the conformal mapping
the area of the hyperbolic triangle $A'$ should not be changed and therefore it reads
\be
S_{A'} = \int\limits_{\triangle A'} |J(z,w)| du dv
\label{eq:trian2}
\ee
where
\be
J(z,w)=\left|\begin{array}{cc} \disp \frac{\partial x}{\partial u}\; & \disp \frac{\partial
y}{\partial u}\; \medskip  \\ \disp \frac{\partial x}{\partial v} & \disp \frac{\partial
y}{\partial v}
\end{array}\right|
\label{eq:jacob1}
\ee
is the Jacobian of conformal mapping $z(w)$. If $z(w)$ is holomorphic function, the Cauchy-Riemann
conditions allow to express $J(w)$ as follows:
\be
J(w) = \left|\frac{dz(w)}{dw}\right|^2
\label{eq:jacob2}
\ee
Let us connect the value of the Jacobian $J(w)$ at the point $w=u+iv$ with the surface height above
the point $(u,v)$ in the plane $w$. Supposing that the surface is defined by the function $f(u,v)$,
we have straightforwardly
\be
J(w) = \sqrt{1+\left(\frac{\partial f}{\partial u}\right)^2+\left(\frac{\partial f}{\partial
v}\right)^2}
\label{eq:surf}
\ee
Hence, the nonlinear partial differential equation
\be
\left(\frac{\partial f(u,v)}{\partial u}\right)^2+\left(\frac{\partial f(u,v)}{\partial v}\right)^2
=\left|\frac{dz(w)}{dw}\right|^4-1
\label{eq:surf2}
\ee
where $z(w)$ is a known function, defines implicitly the height profile $f(u,v)$ of buckling
surface above the plane $w$ tessellated by images of equilateral triangles. By changing $\alpha$ we
can change the effective curvature and study the dependence of the profile on $\alpha$. However,
note that the conformal approach is valid only for $0\le\alpha\le \pi/3$ and fails for
$\alpha>\pi/3$. The function $f(u,v)$ is our desire, which should be later compared with the
wrinkling profile of nonuniformly squeezed elastic membrane.

\subsection{Construction of the conformal map $z(w)$}

The conformal transform $z(w)$ which maps the triangle $M_1M_2M_3$ in $z$ onto the triangle
$M_1'M_2'M_3'$ in $w$ can be realized in two sequential steps as described in \cite{1,4}.

First of all, we map the triangle $M_1M_2M_3$ onto the upper half-plane of auxiliary complex plane
$\zeta=\xi+i\eta$ with three branching points at 0, 1 and $\infty$. This mapping is realized by the
function $z(\zeta)$:
\be
z(\zeta)=\frac{1}{B\left(\frac{1}{3},\frac{1}{3}\right)} \int_0^{\zeta}
\frac{d\zeta'}{\zeta'^{2/3}(1-\zeta')^{2/3}}
\label{eq:z(zeta)}
\ee
where $B\left(\frac{1}{3},\frac{1}{3}\right)$ is the $\beta$-function. The correspondence of the
branching points is as follows:
\be
\begin{array}{lll}
M_1(z=0) & \to & \tilde{M}_1(\zeta=0) \medskip \\
M_2(z=1) & \to & \tilde{M}_1(\zeta=1) \medskip \\
M_3\left(z=e^{-i\frac{\pi}{3}}\right) & \to & \tilde{M}_3(\zeta=\infty)
\end{array}
\label{eq:points1}
\ee

Next step consists in mapping the auxiliary upper half-plane $\im\zeta>0$ onto the circular
triangle $M'_1M'_2M'_3$ with angles $\{\alpha,\alpha,0\}$ -- the fundamental domain of the Hecke
group in $w$. This mapping is realized by the function $\zeta(w)$, which can be constructed as
follows \cite{koppenfels}. Let $w(\zeta)$ be the inverse function of $\zeta(w)$ and represent it as
a quotient
\be
w(\zeta) = \frac{\phi_1(\zeta)}{\phi_2(\zeta)}
\label{eq:w(zeta)}
\ee
where $\phi_{1,2}(\zeta)$ are the fundamental solutions of the 2nd order differential equation of
Picard-Fuchs type:
\be
\zeta(\zeta-1) \phi''(\zeta)+\left(\left(a+b+1\right)\zeta-c\right) \phi'(\zeta)+ab\phi(\zeta)=0
\label{eq:fundam}
\ee
It is known \cite{caratheodory,koppenfels} that the function $w(\zeta)$ conformally maps the
generic circular triangle with angles $\{\alpha_0,\alpha_1,\alpha_\infty\}$ in the upper halfplane
of $w$ onto the upper halfplane of $\zeta$ with the following correspondence of points:
\be
\left\{\begin{array}{ll}
\mbox{In vertex $w(\zeta=0)$} & \mbox{angle $\alpha_0=\pi|c-1|$} \medskip \\
\mbox{In vertex $w(\zeta=1)$} & \mbox{angle $\alpha_1=\pi|a+b-c|$} \medskip \\
\mbox{In vertex $w(\zeta=\infty)$} & \mbox{angle $\alpha_{\infty}=\pi|a-b|$}
\end{array} \right.
\ee
Choosing $\alpha_{\infty}=0$ and $\alpha_0=\alpha_1=\alpha$, we can express the parameters
$(a,b,c)$ of the equation \eq{eq:fundam} in terms of $\alpha$, taking into account that the
triangle $M_1'M_2'M_3'$ in the \fig{fig:02}b is parameterized as follows
$\{\alpha_0,\alpha_1,\alpha_\infty \}=\{\alpha,\alpha,0\}$:
\be
a=b=\frac{\alpha}{\pi}+\frac{1}{2}, \quad c=\frac{\alpha}{\pi}+1
\label{eq:param}
\ee
This leads us to the following particular form of equation \eq{eq:fundam}
\be
\zeta(\zeta-1) \phi''(\zeta)+\left(\frac{\alpha}{\pi}+1\right)(2\zeta-1)
\phi'(\zeta)+\left(\frac{\alpha}{\pi}+\frac{1}{2}\right)^2\phi(\zeta)=0
\label{eq:fundam2}
\ee
where $\alpha = \frac{\pi}{m}$ and $m=3,4,...\infty$. For $\alpha=0$ Eq.\eq{eq:fundam2} takes
especially simple form, known as Legendre hypergeometric equation,
\be
\zeta(\zeta-1) \phi''(\zeta)+(2\zeta-1)\phi'(\zeta)+ \frac{1}{4}\phi(\zeta)=0
\label{eq:fundam3}
\ee
The pair of possible fundamental solutions of \eq{eq:fundam3} reads
\be
\begin{array}{l}
\phi_1(\zeta)=F\left(\frac{1}{2},\frac{1}{2},1,\zeta\right); \medskip \\
\phi_2(\zeta)=iF\left(\frac{1}{2},\frac{1}{2},1,1-\zeta\right)
\end{array}
\label{eq:hyp}
\ee
Substituting \eq{eq:hyp} into \eq{eq:w(zeta)} we get $w(\zeta)$. The inverse function $\zeta(w)$ is
the so-called modular function, $k^2(w)$ (see \cite{modular} for details), i.e.
\be
\zeta(w) \equiv k^2(w) = \frac{\theta_2^4(0,e^{i\pi w})}{\theta_3^4(0,e^{i\pi w})}
\label{eq:zeta(w)}
\ee
where $\theta_2$ and $\theta_3$ are the elliptic Jacobi $\theta$-functions \cite{jacobi}. The
correspondence of branching points for the auxiliary mapping $\zeta(w)$ is as follows:
\be
\begin{array}{lll}
\tilde{M}_1(\zeta=0) & \to & M'_1(w=\infty) \medskip \\
\tilde{M}_2(\zeta=1) & \to & M'_1(w=0) \medskip \\
\tilde{M}_3(\zeta=\infty) & \to & M'_3(w=-1)
\end{array}
\label{eq:points127}
\ee

Collecting \eq{eq:z(zeta)} and \eq{eq:zeta(w)} we arrive at an explicit expression for a composite
map, $z(\zeta(w))$. Computing the Jacobian \eq{eq:jacob2} of this composite conformal mapping, we
get
\be
J(w)\equiv \left|\frac{dz(w)}{dw}\right|^2
=\frac{\left|\eta(w)\right|^8}{\pi^{2/3}B^2(\frac{1}{3},\frac{1}{3})}
\label{eq:ded1}
\ee
where
\be
\eta(w)=e^{\pi i w/12}\prod_{n=0}^{\infty}(1-e^{2\pi i n w}) \quad (w=u+iv)
\label{eq:ded2}
\ee
is $\eta(w)$ the Dedekind $\eta$-function \cite{jacobi}.

The profile $f(u,v)$ of the surface wrinkling above the plane $w=u+iv$ can be computed now from the
Eq.\eq{eq:surf2}, which for the particular choice of the Jacobian $J(w)$, defined by \eq{eq:ded1},
reads
\be
\left(\frac{\partial f(u,v)}{\partial u}\right)^2+\left(\frac{\partial f(u,v)}{\partial v}\right)^2
=\frac{\left|\eta(w)\right|^{16}}{\pi^{4/3}B^4(\frac{1}{3},\frac{1}{3})}-1
\label{eq:surf3}
\ee

\begin{widetext}

\begin{figure}[ht]
\epsfig{file=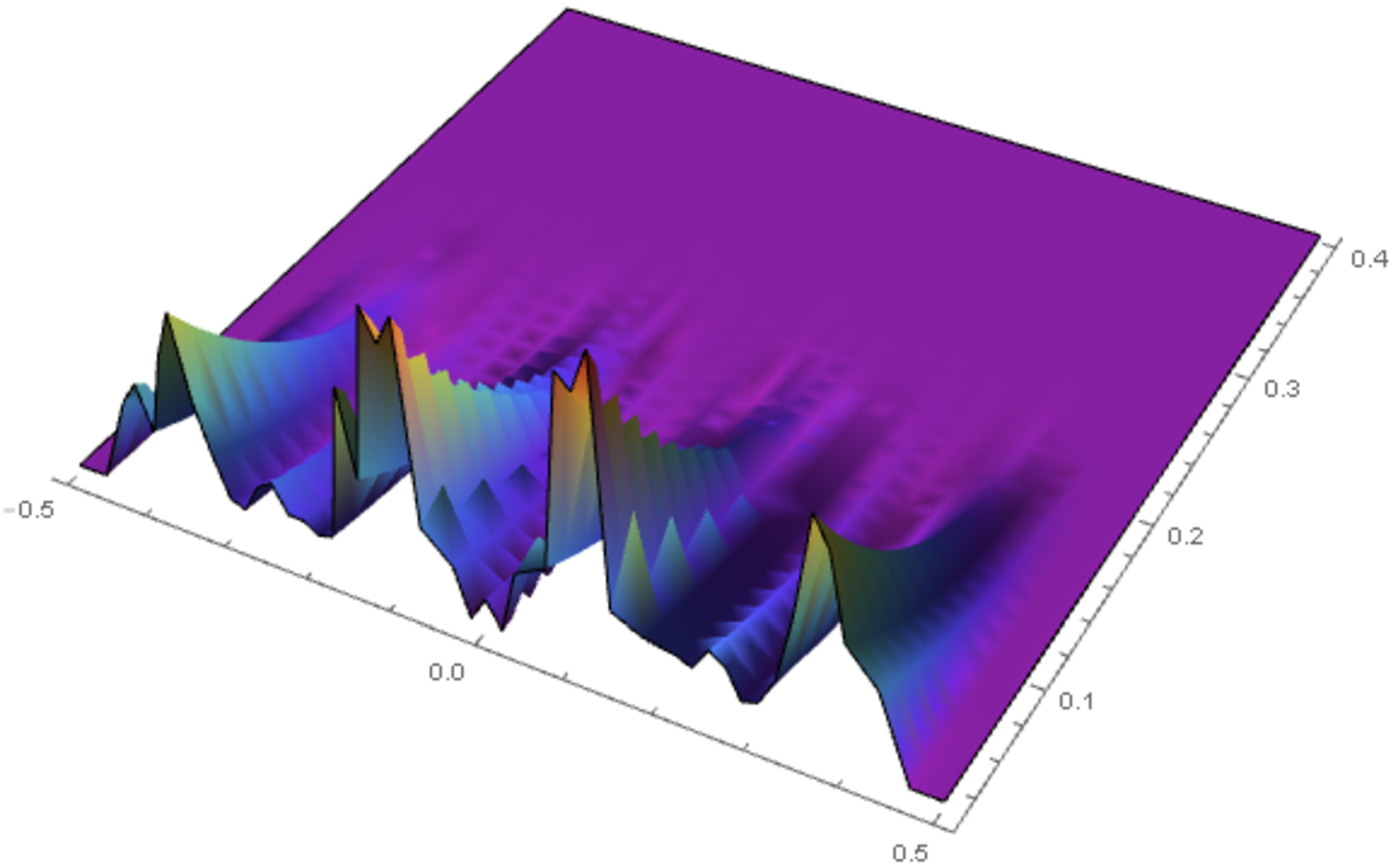,width=7.5cm} \quad \epsfig{file=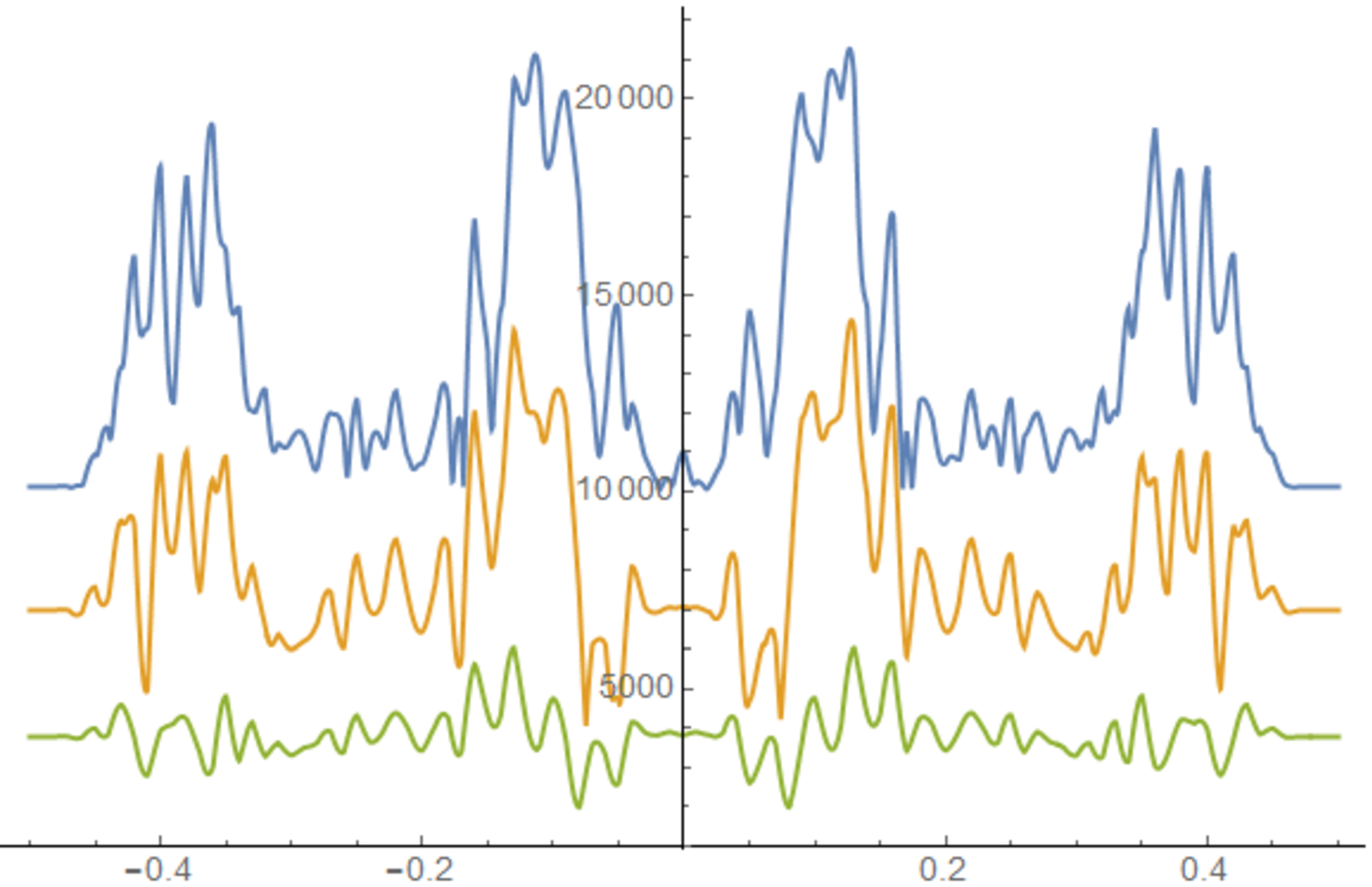,width=7.5cm}
\caption{Numerical solution of the eikonal equation \eq{eq:surf3} for buckling surface in the
geometrical approach (left). Few solutions for the free edge profile in the geometrical approach
for $v=0.07, 0.05, 0.03$ (right).}
\label{fig:02a}
\end{figure}

\end{widetext}

It is worth mentioning that \eq{eq:surf3} has a form of eminent eikonal equation for wave
propagation time to the point $(x,y)$ in the medium that allows for a certain wave speed $c(x,y)$,
namely $|\nabla T(x, y)|^2 = \frac{1}{c(x,y)}$ with the source initially located at some surface
$\Gamma$, with initial Dirichlet conditions \cite{geo1,geo2}. Solving numerically \eq{eq:surf3}, we
get the surface profile $f(u,v)$, shown in the \fig{fig:02a} (left).

Buckling profile near the open boundary $v = 0$, for exponential growth of cells is shown in the
\fig{fig:02a} (right) for few values $v=0.07,0.05,0.03$. Later on we shall compare these profiles
with the ones derived via energetic approach. Note that with progression towards the free edge
(i.e. for $v\to 0$), the shape develops novel peaks vegetating from previous ones. It will be shown
later that the same structure emerges in growth considered in the frameworks of energetic approach.

\section{Energetic approach to buckling}

Here we discuss continuous energetic model of a shape formation in a very thin planar elastic
tissue growing between two impenetrable walls. We conjecture that this model mimics growth in the
situation when cell division works homogeneously with a certain time-dependent rate about the whole
width of the sample squeezed between the walls. Specifically, the growth rate is chosen such that
cell density is gradually increasing towards the direction of growth. Buckling of plants leaves
contribute to the belief, that there exists some correlation length, on which tissue benefits from
local breaking of planar topology, which occurs as a sort of bifurcation and is a fundamental
biological and physical property.

The cell division rate is crucial for resulting profiles, giving rise a variety of buckling forms
and shapes in different biological systems. We put this parameter in our model as the number of
cells, $N(y)$, in a layer $y$, where  $y\equiv y(t)$ characterizes the kinetics of cell division.
The advantage of elastic approach consists in the possibility to develop sufficiently simple
general solution for the stable shape of elastic tissue from the system energy minimization. This
approach can be extended to the model which takes into account random inhomogeneity in growth rate
and in boundary conditions.

\begin{figure}[ht]
\epsfig{file=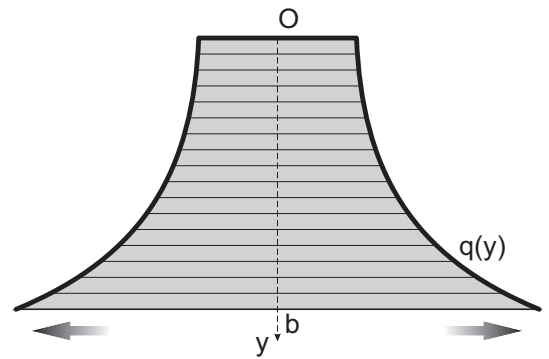,width=7cm}
\caption{Relaxed tissue when two walls have been taken away.}
\label{fig:03}
\end{figure}

The model of the nonuniformly squeezed elastic tissue is set as follows. Consider a 2D sample grown
till time $t$, that buckles out of its plane since it has to grow under the constraints imposed by
walls, that prevent increasing of the in-plane sample width. If we now let the grown tissue to
relax, the tissue will loose its buckling profile and adopt a smooth, not deformed one, as depicted
in the \fig{fig:03}. As long as this resulting state can be taken for the reference one, and we
believe there is no energy loss during the relaxation, the work needed to keep walls is equal to
buckling deformation energy. On next step, we may recreate buckling picture by applying certain
forces to the walls. Eventually, buckling at time $t$ obtained by grown 2D tissue can be understood
in terms of the model of the plate with some sort of gradual prestress applied. Since we consider
linear regime only, there is no in-plane deformations before the plate starts buckling out. Thus
this prestress does not change the size of the plate and the contour length of each row is
conserved as we move from relaxed state to the prestressed one.

Now let us take a closer look on a prestressed plate a thickness $h(a, b\gg h)$ that undergoes
compressing contour forces $q(y)$ distributed along the edges $x = -a/2$ and $x = a/2$, that are
kept fixed. The growth spreads in $y$ direction and the side of the plate $y = 0$, where the
growing starts, is fixed also. The question we will be interested in, is which buckling should
minimize the full energy of the system grown under such conditions? It is known that the energy of
in-plane strain scales as $H_{in} \sim h$ while the energy of out-plane "buckling-kind"
deformations scales as $H_{out} \sim h^3$, which implies that thin enough plates would benefit from
buckling instead of staying stable. It will be shown further that buckling complexity crucially
depends on the thickness.

In tense state the full energy, $H$, is the sum of deformation energy, $U$, and of its potential
counterpart of applied forces, $P$, i.e. $H=U+P$. By minimizing the energy functional, the equation
on equilibrium buckling profile can be obtained. The main steps of the derivation are presented
here, while more elaborated description constitutes the subject of the Appendix A.

In linear approximation the deformation energy of an elastic sample can be written in terms of
tension and deformation tensor components:
\be
U = \frac{1}{2}\int\limits_V \bigg(\sum_i\sigma_i\epsilon_i +\sum_{i\neq j}
\tau_{ij}\gamma_{ij}\bigg) dV
\label{1}
\ee
where $i=\{x,y,z\}, j=\{x,y,z\}$, $i\neq j$, and $\tau_{ij}=\tau_{ji}$, $\gamma_{ij}=\gamma_{ji}$.

The potential energy of applied forces is measured by the work needed to get from this state to the
reference one which is regarded as an unbuckling state.
\be
P = \int\limits_{-a/2}^{a/2} \int\limits_{0}^{b} q(y)\left[w\partial^2_{xx} w + \left(\partial_x
w\right)^2\right] dxdy,
\label{2}
\ee

Minimizing the resulting energy functional, we get the Euler equation for the equilibrium shape
$w(x,y)$
\be
D\Delta^2 w(x,y) + P = 0
\label{3}
\ee
where $D = \frac{Eh^3}{12(1 - \mu^2)}$ is bending stiffness of a plate, $P = 3q(y)\frac{\partial^2
w}{\partial x^2}$, and $\Delta = \partial_{xx}^2+\partial_{yy}^2$.

Equation \eq{3} is a linear form of the well-known F\"oppl-von-Karman (FvK) equation
\cite{muller,lewicka,gemmer} for out-of-plane displacement of a thin homogeneous rectangular plate.
It can give us some insight on the eigenfunctions of the plate, however due to its linear
structure, \eq{3} can say nothing neither about dynamics of the system on the phase plane
"force-harmonics" between different bifurcation points, nor about the types of these bifurcations.
The nonlinear parts should be analyzed to understand the process deeper. However external errors,
for example at the boundaries, can increase the number of intersections of different trajectories
on the phase plane and effectively smear some of nonlinear peculiarities. As a result, the system
may exist in many eigenstates at the same time and the problem of choosing between different states
becomes purely statistical.

To simplify forthcoming espressions, rewrite \eq{3} as:
\be
D\left(\partial^4_{xxxx} w + 2\partial^4_{xxyy} w + \partial^4_{yyyy} w\right) +
q(y)\partial^2_{xx} w = 0
\label{4}
\ee
where the reassignment $q \to 3q$ is made.

Note that in the limit of infinite growing times and constant compressing force, we arrive at the
Euler equation for rod bending. The solution for eigenfunctions takes the well-known form, $w_n(x)
= A_n \sin\frac{n\pi x}{a}$, where $q_n = \frac{n^2 \pi^2 D}{a}$ are the eigenvalues of
corresponding Sturm-Liouville problem.

Our growth model implies that the plate is fixed on three sides, namely at $x = -\frac{a}{2}$, $x=
\frac{a}{2}$ and $y = 0$, where the $z$-displacement, $w(x,y)$ nullifies. Moreover deformations
nullify as well since our plate can freely rotate at the  boundaries. At forth open edge (the
surface) both, tension and forces, vanish (see Appendix B for details):
\be
\left\{\begin{array}{r}
w\left(-\frac{a}{2}, y\right)  =  \partial^2_{xx} w\left(-\frac{a}{2}, y\right) = 0 \medskip \\
w\left(\frac{a}{2}, y\right) = \partial^2_{xx} w\left(\frac{a}{2}, y\right) = 0 \medskip \\
w(x, 0) = \partial^2_{yy} w(x, 0) = 0 \medskip \\
\partial^2_{yy} w(x, b) + \mu\partial^2_{xx} w(x, b) = 0 \medskip \\
\partial^3_{xxx} w(x, b) + (2-\mu)\partial^3_{xxy} w(x, b) = 0
\end{array}\right.
\label{bound}
\ee

The symmetry of the problem forces the solution to be even or odd in $x$ coordinate (relatively to
the center of the plate). Thus, we seek the solution in terms of sine- or cosine-Fourier series in
$x$. The resulting coefficients are weighted functions of $y$. To satisfy boundary conditions at $x
=\frac{a}{2}$ and $x= -\frac{a}{2}$, one should require the sine series to be over only even
harmonics and the cosine series - over odd harmonics:
\be
\begin{array}{l}
\disp w_S(x, y) = \sum_{n = 0}^{\infty} Y_n(y) \sin \frac{2\pi n x}{a} \medskip \\
\disp w_C(x, y) = \sum_{n = 0}^{\infty} Y_n(y) \cos \frac{(2n+1)\pi x}{a}
\end{array}
\label{fourier}
\ee

Substituting \eq{fourier} in \eq{4}, we get the following fourth-order equation for $Y_n(y)$:
\be
\partial^4_{yyyy} Y_n(y) - 2 S_n \partial^2_{yy}Y_n(y) + S_n \left(S_n -
\frac{q(y)}{D}\right)Y_n(y) = 0
\label{y(n)}
\ee
where
\be
S_n = \left\{\begin{array}{ll} \left(\frac{2n \pi}{a}\right)^2 & \mbox{for sine-solutions}
\medskip \\
\left(\frac{(2n+1) \pi}{a}\right)^2 & \mbox{for cosine-solutions} \end{array} \right.
\label{Sn}
\ee

Equations \eq{y(n)}-\eq{Sn} are equipped with the following boundary conditions:
\be
\left\{\begin{array}{r}
Y_n(0) = \partial^2_{yy}Y_n(0) = 0 \medskip \\
\partial^2_{yy}Y_n(b) - \mu S_n Y_n(b) = 0 \medskip \\
\partial^3_{yyy}Y_n(b) - (2-\mu) S_n \partial_{y}Y_n(b) = 0
\end{array} \right.
\label{bound2}
\ee

The Sturm-Liouville problem \eq{y(n)}-\eq{bound2} cannot be solved analytically for generic
squeezing protocol $q$, however can be treated exactly for $q(y) = q ={\rm const}$, since in that
case it becomes a linear differential equation with constant coefficients. The solution for
$q(y)=q$ reads
\begin{multline}
Y_n(y) = A_1 \sin\alpha_n y + A_2 \cos \alpha_n y  \\
+ A_3 \sinh \beta_n y + A_4 \cosh \beta_n y,
\label{sol-y}
\end{multline}
where
\be
\alpha_n = \sqrt{S_n}\sqrt{-1 + \sqrt{\frac{q}{D} S_n}}; \quad
\beta_n = \sqrt{S_n}\sqrt{1 + \sqrt{\frac{q}{D} S_n}}
\label{eig-y}
\ee
Taking into account boundary conditions \eq{bound2}, we obtain:
\begin{itemize}
\item The amplitude,
\be
Y_n(y) \sim \sinh \beta_n y + \frac{\beta_n^2 - \mu S_n}{\alpha_n^2
+\mu S_n}\frac{\sinh \beta_n b}{\sin \alpha_n b} \sin \alpha_n y
\label{amp}
\ee
\item The spectral equation, which eventually determines the plate buckling modes,
\be
\frac{\alpha_n}{\beta_n} \tanh {\beta_n b} \cot {\alpha_n b} \frac{\beta_n^2 -
\mu S_n}{\alpha_n^2 + \mu S_n} \frac{\alpha_n^2 - (2-\mu)
S_n}{\beta_n^2 - (2-\mu) S_n} = 1
\label{modes}
\ee
\end{itemize}

\begin{widetext}

\begin{figure}[ht]
\epsfig{file=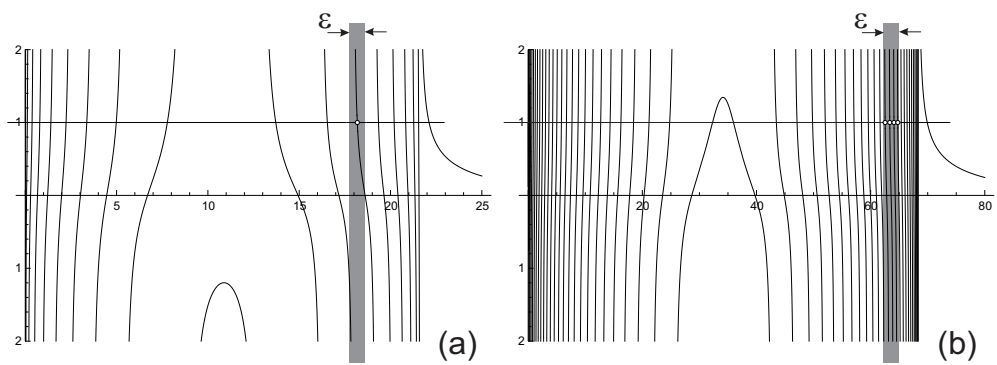,width=13cm}
\caption{Graphic solution of \eq{modes} for the plate with $a = 5$, $b = 2$ and (a)
$\frac{q}{D} = 100$; (b) $\frac{q}{D}=1000$.}
\label{fig:05}
\end{figure}

\end{widetext}

Eq.\eq{modes} allows to determine harmonics, $n$. It can be inferred from the \fig{fig:05}a that
when the effective compressing parameter $\frac{q}{D}$ is small enough, only few harmonics are
available in the plate. By increasing $\frac{q}{D}$ we increase number of possible harmonics, as
shown in the \fig{fig:05}b. Let us argue that in any real physical system, the unavoidable presence
of noise makes the buckling "uncertain", meaning simultaneous presence of many harmonics,
corresponding to the solution of \eq{modes}. Indeed, suppose that due to fluctuations in initial
conditions, the solution of \eq{modes} is known with some accuracy, $\epsilon$, designated in the
\fig{fig:05}a,b by the grey strip of width $\epsilon$. If $\epsilon$ is smaller than spacing
between harmonics, only one solution is realized as shown in the \fig{fig:05}a. Otherwise, one
could have many solutions inside the strip of width $\epsilon$ -- see \fig{fig:05}b.

As approved by \fig{fig:05}b, for soft enough surfaces buckling is mostly pronounced and the
spectrum should be extremely dense. In this case, the gap between the energies of neighbouring
harmonics (local minimizers) is much less the energy perturbations, induced by the errors. As long
so, one can imply that the system lives in all the states at one time. In other words, the
probabilities of the each eigenvector is the same. This allows to take all the weight functions in
Fourier series with equal absolute amplitudes. But still the sign of the each wave amplitude is a
subject that provides this problem with stochastic nature, because states that differ only by a
sign have equal probabilities to survive.

\begin{widetext}

\begin{figure}[ht]
\epsfig{file=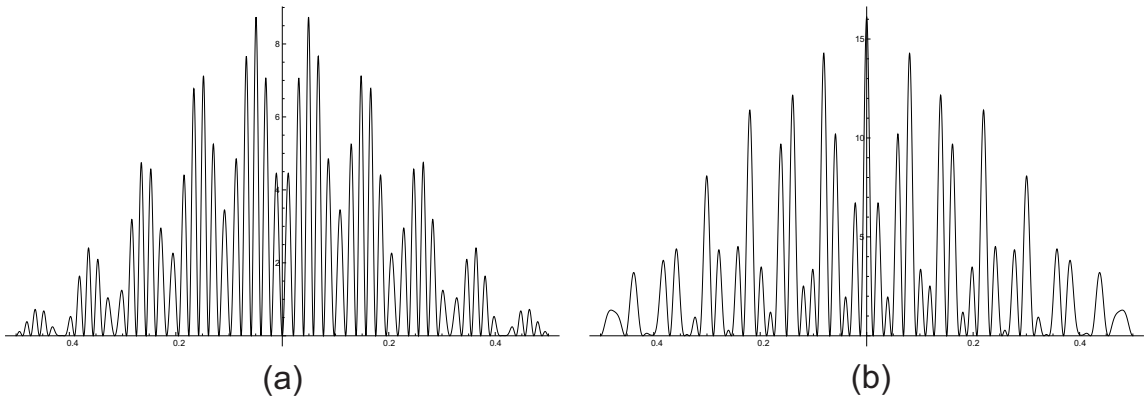,width=15cm}
\caption{(a) Even and (b) odd buckling profiles for plates with $a = 5$, $b = 2$ and
$\frac{q}{D} = 10^3$.}
\label{fig:7}
\end{figure}

\end{widetext}

We plot the squared buckling profile at the free edge, $w^2(x,b)$, in \fig{fig:7} for even and odd
solutions. Interestingly, the profiles seem very similar, being different in small details only,
such as a peak in the center. It can be seen from the \fig{fig:7}, that constant compressing
protocol gives homogeneous buckling with no fractal structure pronounced at the free edge.
Nevertheless, homogenous squeezing is still of great importance since it gives us physical insight
on the spectrum properties of the deformed plate.

Now we are going to proceed with numerical approach for non-constant $q(y)$ looking for fractal
behavior. It has to be mentioned that in our approach we deal only with thin plates. It is well
justified because all interesting features, such as intense buckling and fractal profiles, can be
found only in soft enough systems, which means low values of bending stiffness, $D$. Thus having
the thickness increased and jumping to 3D problem we leave little chances for buckling, according
to spectral equation \eq{modes}.

\section{Numerical solution of linear problem}

For nonconstant squeezing protocol, $q(y)$, one solves \eq{y(n)} numerically for various modes $n$
and receive the $y$-dependent amplitudes for each harmonics. To proceed, replace each derivative in
\eq{y(n)} by its finite-difference analog and convert it along with the
boundary conditions \eq{bound2} into the system of linear equations for the vector $u$:
\be
A_N{\bf u} = {\bf b},
\label{A}
\ee
where $N$ is the number of steps in the finite-difference scheme that divides the mattress length
$a$ in equal step sizes $h = a/N$.

The matrix $A_N$ is five-diagonal and approximates the equation \eq{y(n)} with the accuracy
$O(h^2)$. Details of the derivations approximations and the explicit forms of matrix elements can
be found in Appendix C. According to \eq{y(n)}-\eq{bound2}, the row ${\bf b}$ should be a null
vector. In this case, we get only trivial solution ${\bf u} = 0$ for any nonsingular $A_N$. It is
well justified that semi-analytical spectral condition for non-constant compressing, analogous to
\eq{modes}, is
\be
\det A_N(n) = 0
\label{det}
\ee
for any large $N$.

Unfortunately, Eq.\eq{det} is difficult to analyze due to general computational reasons, like
increasing errors for sufficiently small step sizes, $h$. However, in linear regime one can legally
find order of magnitude of critical forces $q(y)$ when first bucking appears. It is worth noting
that the spontaneous symmetry breaking at the first buckling point, crucially depends on noise at
boundaries of the plate and fluctuations in the bulk. Thus, in order to describe buckling profile
we treat $b$ as a noise of small amplitude. Fluctuations in the lattice give rise to random
matrices implementation. On this way one can improve \eq{A} and write $M_N = A_N + R_N$, where
$A_N$ is the matrix described above and $R_N$ is a random $N \times N$ matrix describing the
perturbation. Then improved solution of \eq{A} takes the following form:
\be
{\bf u} = {\bf u}_A + \sum {\bf u}_i P(R_i),
\label{u}
\ee
where ${\bf u}_A$ is not perturbed solution with the matrix $A$, and ${\bf u}_i$ is the "random
solution" with the matrix $R_i$ such that $R_i {\bf u}_i = {\bf b}$, and $P(R_i)$ is the
probability of the fluctuation $R_i$.

\begin{widetext}

\begin{figure}[ht]
\epsfig{file=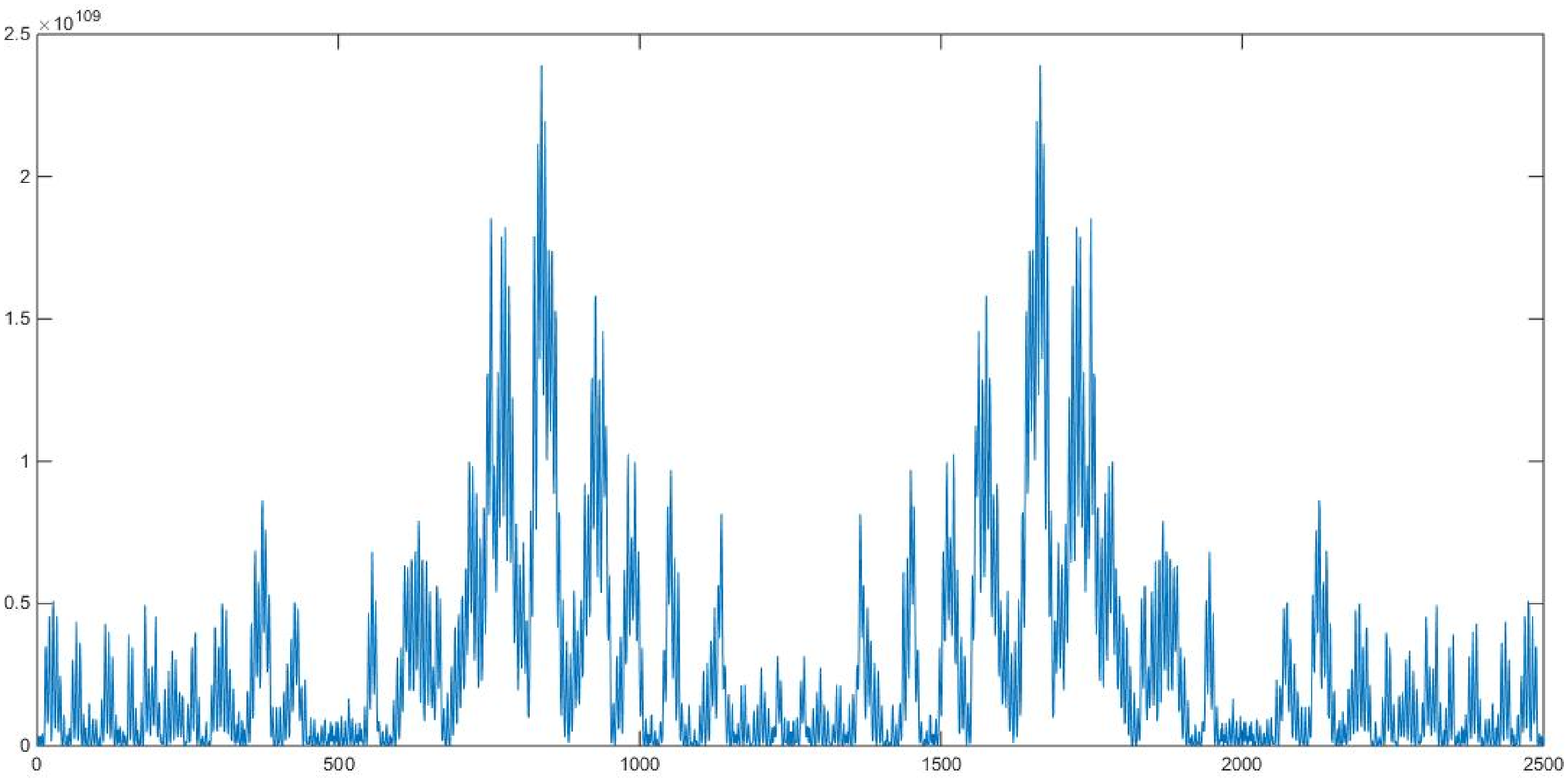, width = 8cm} \epsfig{file=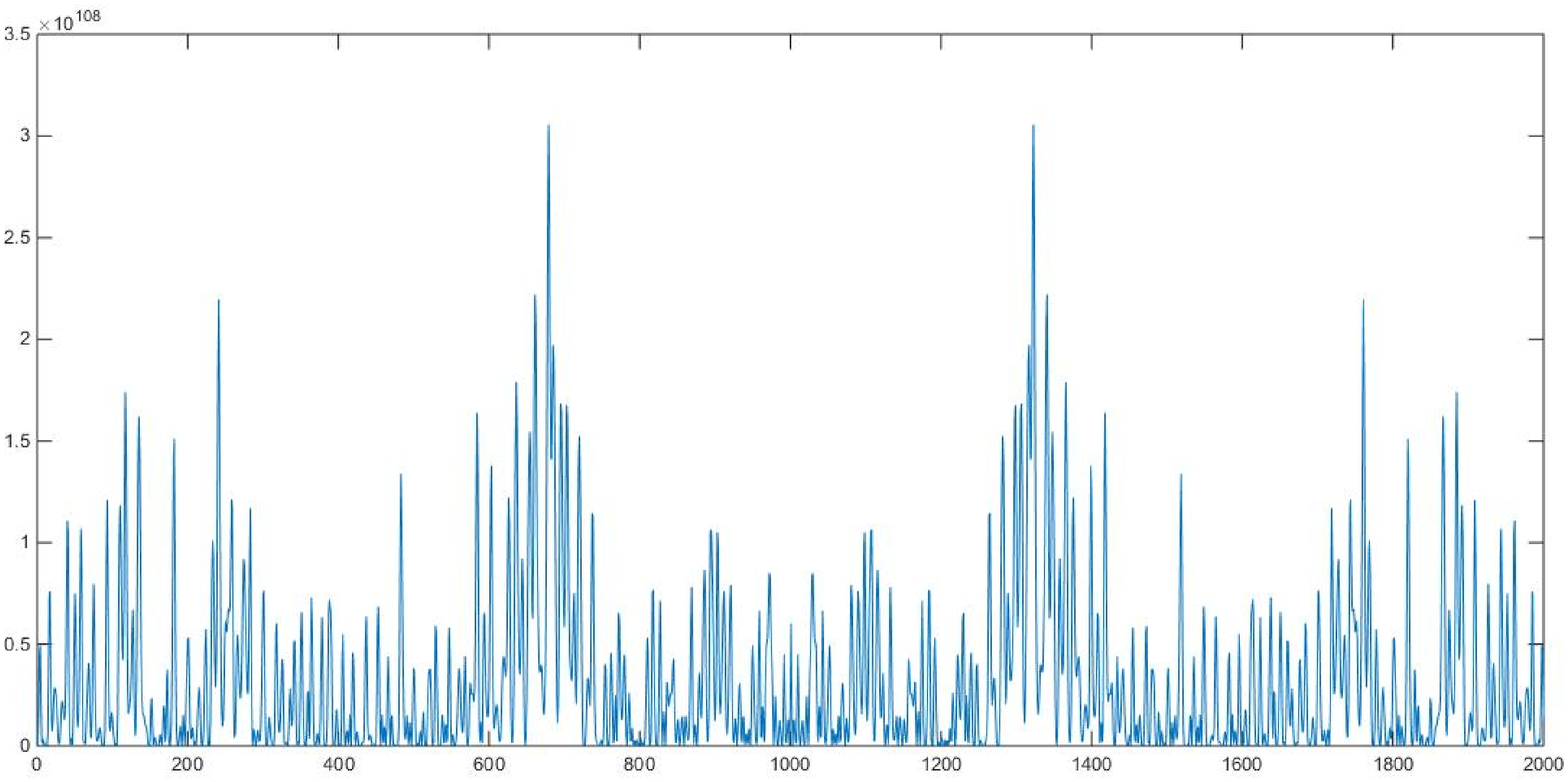, width = 8cm}
\caption{(left) Even buckling profile $w^2(x, b)$ on the free edge for the plate with $a = 5$,
$b = 1$ for $q(y) \sim e^y$; (right) Odd buckling profile $w^2(x, b)$ on the free edge for the
plate with $a = 10$, $b = 1$ for $q(y) \sim e^y$.}
\label{fig:8}
\end{figure}

\end{widetext}

Discussion of specific properties of random matrices $R$ goes beyond the scopes of this work. We
assume that it is possible to account for noise taking ${\bf b}$ as small random perturbation.
First and last two elements of ${\bf b}$ describe noise at edges and other elements are responsible
for fluctuations in the bulk of the plate. Solving \eq{A} numerically with such choice of ${\bf
b}$, one finds typical buckling profiles.

Let us now look on odd and even profiles on the free edge for exponential compressing protocol
$q(y) \sim e^y$ \fig{fig:8}. First, it can be noticed that at the center of the plate the fractal
behavior qualitatively does not change. It was found that for larger $\frac{a}{b}$ new higher modes
develop in the plate whose relative disposition mimics the picture at low $\frac{a}{b}$ and
positions of peaks on subsequent hierarchical levels. These pictures were found stable to errors
induced by the boundaries, especially at the central part of the plate. These profiles have
striking similarities with the profiles derived via geometric approach, \fig{fig:02a}.
Interestingly, no such self-similarity profiles have been found for other types of compressing
protocol, $q(y)$, including the case of constant squeezing.

\begin{figure}[ht]
\epsfig{file=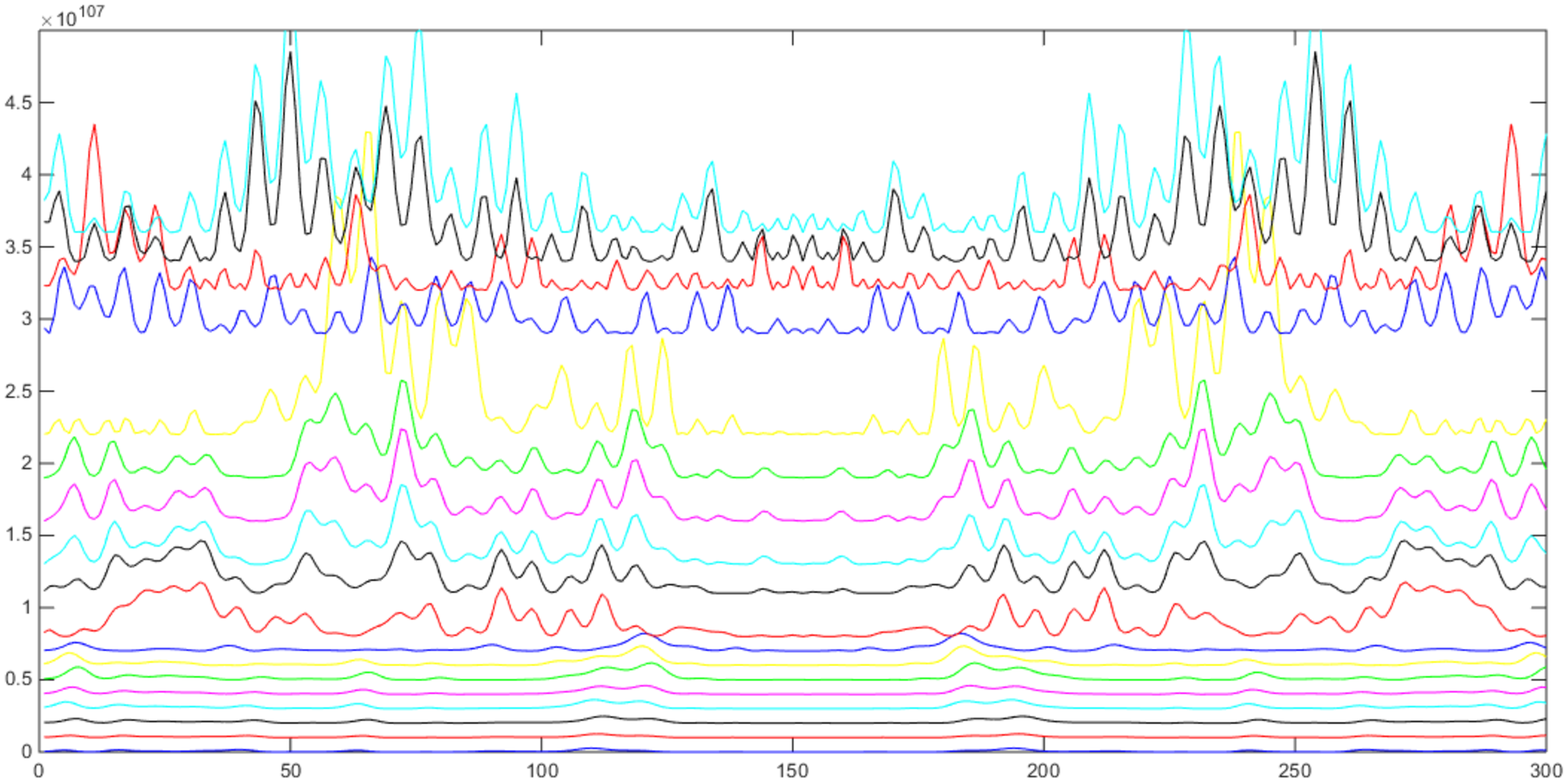, width=9cm, height = 7cm}
\caption{The developing of new modes with the progression towards the free edge}
\label{fig:12}
\end{figure}

To show the effect of developing branching on the buckling surface with progression of $y$ towards
the free edge, we have plotted the squared profiles, $w(x,y)$ on $x$ for different $y$, starting
from $y = 0$, see \fig{fig:12}. Developing of new peaks as a result of fission of the previous ones
is consistent with the model of cell division, as long as we used exponential squeezing protocol
$q(y)$ in our computations.

\section{Conclusion}

In this paper,  using geometrical and energetic approaches, we have investigated equilibrium shapes
of exponentially growing 2D patterns squeezed between plate boundaries. The resulting profiles
posses self-similar organization, one of most crucial features of exponential protocol of growth.
This can explain visual resemblance of various biological structures in the nature and gives rise
for mechanical description of cells proliferation.

The parameters of the system, $h, E$ and $\mu$, can be combined into one effective stiffness
$D=\frac{Eh^3}{12(1 - \mu^2)}$ of the plate as the main parameter defining mechanical properties of
the system. It has been shown for constant compressing, that increasing stiffness suppresses
buckling. Thus as long as $h$ provides the strongest influence on $D$, it seems meaningless to look
for buckling in thick stiff membranes. At the same time, thick but soft membranes can be
reformulated in terms of thin plates if one rescales the thickness and Young modulus to keep the
effective stiffness $D$ the same.

We can provide physical arguments in favor of multi-solution property of spectral equation
\eq{modes}. As long as \eq{modes} is derived by the total energy minimization, each mode allowed by
\eq{modes} realizes a local minimum of the sample's energy. When the plate is rigid enough, only
one single mode exists, and corresponds to the energy minimum that is surely the global one.
However, for rather soft plate, we have many energy levels separated by small gaps. It is known
that global minimizer of the F\"opple-von-Karman elastic energy of a thin plate is not that which
is found in real experiments \cite{klein1, klein2}, where multi-waves solution were encountered
instead. Moreover, with decrease of the thickness, the number of waves is increasing and resulting
profile is getting more buckled, according to the mentioned experiments. This conclusion is in good
agreement with \fig{fig:05}, which can be regarded (since as $D \sim h^3$) as cases of two
different thicknesses. Apparently, due to fluctuations in the system, the profile of the plate can
be understood as the superposition of waves (local minimizers) taken with certain magnitudes
accordingly to Boltzmann probabilities.

Let us finalize the paper by some observation which is, to our point of view, beyond just a funny
speculation. It is well known that a thin long rod, like a dry noodle, being squeezed at
extremities, breaks typically in three parts at two symmetric points to the center \cite{ignobel}
(this work has earned the Ig Nobel prize in 2006). Two maxima of largest amplitudes on the $w(x)$
curve, clearly seen in the \fig{fig:8} appeared in any of our solutions, in geometric and energetic
approaches, in odd or even cases, for many different compressing protocols $q(y)$. It seems
naturally to conjecture that points of largest amplitudes in $w(x)$ coincide with the points of
maximal stress in the plate. So, we could conjecture that a nonuniformly squeezed plate breaks by
proliferation of two basic cracks started at the free surface of points of maximal stress.

\begin{acknowledgments}

We are grateful to L. Mirny and M. Tamm for fruitful discussions and A. Orlov for the valuable
expertise in numerical approach. S.N was partially supported by the IRSES DIONICOS grant.

\end{acknowledgments}

\begin{appendix}

\section{Minimizing full energy of a plate}

In tense state the full energy is determined by the deformation energy and its potential
counterpart of distributed forces, $H = U + P$. In the linear approximation the deformation energy
of a sample can be written in terms of tension and deformation tensor components (see \eq{1}):
\be
U = \frac{1}{2}\int_V \bigg(\sum_i\sigma_i\epsilon_i +\sum_{i\neq j} \tau_{ij}\gamma_{ij}\bigg) dV
\label{app:1}
\ee
where $i=\{x,y,z\}, j=\{x,y,z\}$, $i\neq j$, and $\tau_{ij}=\tau_{ji}$, $\gamma_{ij}=\gamma_{ji}$.

Proceeding with Hooke's law and taking into account the definition of the Young modulus $E$ and the
Poisson ratio $\mu$, we get the following explicit expressions:
\be
\begin{array}{ll}
\disp \epsilon_x = \frac{\sigma_x - \mu\sigma_y - \mu\sigma_z}{E}; & \disp \gamma_{xy} = \frac{2(1
+ \mu)}{E}\tau_{xy} \medskip \\
\disp \epsilon_y = \frac{\sigma_y - \mu\sigma_x - \mu\sigma_z}{E}; & \disp \gamma_{xz} = \frac{2(1
+ \mu)}{E}\tau_{xz} \medskip \\
\disp \epsilon_z = \frac{\sigma_z - \mu\sigma_x - \mu\sigma_y}{E}; & \disp \gamma_{yz} = \frac{2(1
+ \mu)}{E}\tau_{yz}
\end{array}
\label{app:2}
\ee

Then one can reverse \eq{app:2} and substitute the result into the deformation energy
\eq{app:1}. For in-plane compression just until the bifurcation starts, all $z$--components of
deformation and stress tensors vanish: $\sigma_z$ = $\tau_xz$ = $\tau_yz$ = 0. It is also worth
mentioning that all deformations are considered in a midplane of our thin plate. Thus, we get:
\be
U = \frac{E}{2(1 - \mu^2)}\int_V \left(\epsilon_x^2 + 2\mu\epsilon_x\epsilon_y + \epsilon_y^2 +
\frac{1 - \mu}{2}\gamma_{xy}^2\right) dV
\label{app:3}
\ee

In linear theory it is possible to treat the components of deformations in terms of linear
dependencies of in-plane displacements:
$$
\epsilon_x = \partial_x u, \quad \epsilon_y = \partial_y v \quad \gamma_{xy} = \partial_y u +
\partial_x v
$$
where $u(x,y)$ and $v(x,y)$ are in-plane displacements in $x$ and $y$ directions of the plate
correspondingly.

It should be mentioned that we do not take into account deformations due to contraction, as usually
is done in pre-buckling state. This effect introduces changes in deformation of second order of
$\frac{h}{a}$ and thus could be neglected in the linear approximation. Designate the transverse
displacement of midplane points in the instable state as $w(x,y)$ and see how bending (i.e.
buckling) induces the local deformations $u_{\rm b}$ and $v_{\bf b}$ at some height $z$ above the
midplane. Under the assumption for thin plates that the normal is the same for all the planes of the plate
and slightly change the direction, we can write:

\be
u = -z\partial_x w, \quad v = -z\partial_y w
\label{app:4}
\ee

Now it is easy to write down the deformation energy in terms of displacement, $w(x,y)$. Integrating
over the thickness of the plate, one arrives at the following expression:
\begin{multline}
U = \frac{D}{2}\int_{-a/2}^{a/2}\int_{-b/2}^{b/2} \Big\{\left(\partial^2_{xx} w + \partial^2_{yy}
w\right)^2 \\ +2(1 -\mu)\left(\left(\partial^2_{xy} w\right)^2 - \partial^2_{xx} w \partial^2_{yy}
w\right) \Big\}dxdy
\label{app:5}
\end{multline}
where $D = \frac{Eh^3}{12(1 - \mu^2)}$ is called a bending stiffness of the plate.

The potential energy of applied forces can be treated as the work needed to get from this state to
the reference (unbuckling) state. Let $p(y)$ be the pressure applied at borders of the plate. It
can be easily converted to contour force density, $q(y) = h p(y)$, since applied forces are uniform
along the plate thickness. The increment of a total work is a sum of a work on $u$ displacement
tangential to the plate, and work on $w$ displacement (the force does not perform any work on the
displacements normal to the $x$-axis). The tangential part of local force is $dF_{\rm tan} =
\frac{p(y)dydz}{\partial_y w}$ and the part that performs work on z-displacements is $dF_z =
p(y)\partial_x w\, dydz$. Using the relations \eq{app:4} one can write the potential energy as the
whole work performed:
\begin{multline}
P = \int_{0}^{b}\int_{-h/2}^{h/2} \left(-\frac{z}{\partial_y w} + w\right) p(y) \partial_x w\, dydz
\\ = h\int_{0}^{b} p(y)w \partial_x w\, dy,
\label{app:6}
\end{multline}
where we have integrated over the thickness and the work on tangential displacements vanishes due
to effectively zero compression energy in the linear approximation. Rewrite \eq{app:6} as:
\be
P = \int\limits_{0}^{b} q(y)\left(w\partial^2_{xx} w + \left(\partial_x w\right)^2\right) dxdy,
\label{app:7}
\ee

Combining \eq{app:5} and \eq{app:7} in one energy functional to be minimized, write down the
corresponding Euler equation:
\begin{multline}
\frac{D}{2}\Big\{\partial^2_{xx}\left(2\left(\partial^2_{xx} w +
\partial^2_{yy} w\right) - 2(1 - \mu)\partial^2_{yy} w\right) \\
+\partial^2_{yy}\left(2\left(\partial^2_{xx} w + \partial^2_{yy}
w\right) - 2(1 - \mu)\partial^2_{xx} w\right) \\
+2(1 -\mu)\partial^2_{xy}\left(2\partial^2_{xy} w\right)\Big\} + 3q(y)\partial^2_{xx} w= 0
\end{multline}
and eventually receive a biharmonic equation on $w(x, y)$
\begin{equation}
D\Delta^2 w + P = 0
\end{equation}
where $P = 3q(y)\partial^2_{xx} w$

\section{Derivation of boundary conditions}

To derive boundary conditions we should first discriminate between two different types of fixation.
If the plate is firmly fixed at a border and the rotation over the line of fixing is prohibited,
than it is clamp type of fixation and the value $\partial_{\bf n} w$ should vanish, where ${\bf n}$
is a normal towards the border. Otherwise rotation is allowed meaning joint type of fixation. In
the last case we demand the absence of any deformations at the edges.

Taking into account the dependencies between bending function $w(x, y)$ and $u-v$ displacements,
one can find that the condition of absence of the deformation on $x = -a/2, x = a/2$ edges is
equivalent to $\partial^2_{xx} w = 0$. Similar condition is true for $y = 0$ edge.

In linear elasticity regime there are no deformations along the plate in pre-bifurcation states. It
means that sizes of a plate in $x$ and $y$ dimensions are kept constant. Thus the 2nd boundary
condition says that $z$-displacement nullifies at the boundary, i.e. $w = 0$.

While the three edges of a rectangular plate are fixed, the last one, at $y = b$ is kept free to
bend. The natural condition here is vanishing of forces and tensions in the direction of growing.
Let us write down the Hooke's law that describes the dependencies between tension and deformation
tensors components in the linear regime (compare to \eq{app:2}):
\be
\begin{array}{rcl}
\disp \sigma_x = \disp  \frac{E}{1 - \mu^2} (\epsilon_x + \mu\epsilon_y) & = & \disp \frac{E}{1 -
\mu^2} (\partial_x u + \mu\partial_y v); \medskip \\
\disp \sigma_y = \disp \frac{E}{1 - \mu^2} (\mu\epsilon_x + \epsilon_y) & = & \disp \frac{E}{1 -
\mu^2}(\mu\partial_x u + \partial_y v); \medskip \\
\disp \tau_{xy} = \disp \frac{E}{2(1 + \mu)}\gamma_{xy} & = & \disp \frac{E}{2(1 + \mu)}
\left(\partial_y u + \partial_x v\right)
\end{array}
\label{bc}
\ee

According to \eq{bc}, to vanish tensions towards $y$ axis one should require $\partial_{yy} w +
\mu\partial_{xx} w = 0$. The same condition for forces can be derived easily, because we take the
thickness $h$ of the plate doesn't change through bending. The change of forces per length unit
projection on $y$-axis $df_y = \frac{\delta F_y}{\delta x}$ acting on the platform with sizes
$\delta y,\delta x, h$ near the free edge should be zero:
\begin{multline}
\delta (df_y) = h(\sigma_y(y + dy) - \sigma_y(y)) + h(\tau_{xy}(y + dy) - \tau_{xy}(y)) \\
= h \delta y (\partial_y \sigma_y +\partial_y \tau_{xy}) \\ = h \delta y
\frac{E}{1-\mu^2}\left(\partial_{yyy} w + (2 - \mu)\partial_{xxy} w \right)
\end{multline}
This gives us the second boundary condition at $y = b$:
\be
\partial_{yyy} w + (2 - \mu)\partial_{xxy} w = 0
\ee

\section{Numeric approach}

To approximate \eq{y(n)} we use symmetric difference quotient with the precision $O(h^2)$:
\begin{multline}
\frac {u_{i-2} - 4u_{i-1} + 6u_{i} - 4u_{i+1} + u_{i+2}}{h^4} \\
-2S_n \frac{u_{i-1} - 2u_{i} + u_{i+1}}{h^2} + S_n\left(S_n - \frac{q_i}{D}\right)u_{i}=0
\label{approx}
\end{multline}
for $2\le i\le N-2$, where $S_n = \left(\frac{2\pi n}{a}\right)^2$ or $S_n =
\left(\frac{(2n+1)\pi}{a}\right)^2$ and $h =\frac{a}{N}$ is a step size.

The boundary conditions are approximated by right-hand and left-hand quotients (to not to introduce
additional points outside the plate). To have precision $O(h^2)$ they require one more point:
\be
\begin{array}{l}
u_N' =  \disp \frac {3u_{N} + u_{N-2} - 4u_{N-1}}{2h} \medskip \\
u_0'' = \disp \frac {2u_{0} - 5u_{1} + 4u_{2} - u_{3}}{h^2} \medskip \\
u_N'' = \disp \frac {2u_{N} - 5u_{N-1} + 4u_{N-2} - u_{N-3}}{h^2} \medskip \\
u_N''' = \disp \frac {5u_{N} - 18u_{N-1} + 24u_{N-2} - 14u_{N-3} + 3u_{N-4}}{2h^3}
\end{array}
\ee

We can rewrite \eq{approx} using 5-diagonal matrix $A$ having the following form:

\begin{widetext}
\be
A = \frac{1}{h^2}\left( \begin{matrix}
h^2 & 0 & 0 & 0 & 0 & 0 & 0 & \cdots & 0 \medskip \\
2 & -5 & 4 & -1 & 0 & 0 & 0 & \cdots & 0 \medskip \\
a_{0}^2 & a_{1}^2 & a_{2}^2 & a_{3}^2 & a_{4}^2 & 0 & 0 & \cdots & 0 \medskip \\
0 & a_{1}^3 & a_{2}^3 & a_{3}^3 & a_{4}^3 & a_{5}^3 & 0 & \cdots & 0 \medskip \\
0 & 0 & a_{2}^4 & a_{3}^4 & a_{4}^4 & a_{5}^4 & a_{6}^4 & \cdots & 0 \medskip \\
 &  & & \ddots & \ddots & \ddots & \ddots & \ddots  \medskip \\
0 & 0 & \cdots & 0 & a_{N-4}^{N-2} & a_{N-3}^{N-2} & a_{N-2}^{N-2} & a_{N-1}^{N-2} &
a_{N}^{N-2} \medskip \\
0 & 0 & \cdots & 0 & 0 & -1 & 4 & -5 & 2 - \mu h^2 S_n \medskip \\
0 & 0 & \cdots & 0 & a_{N-4}^{N} & a_{N-3}^{N} & a_{N-2}^{N} & a_{N-1}^{N} & a_{N}^{N}
\end{matrix}
\right)
\ee
\end{widetext}

where we have introduced the definitions for $2\le i\le N-2$
\be
\begin{array}{l}
a_{i}^{i}  =  \disp \frac {6}{h^2} + 4S_n + h^2 S_n\left(S_n - \frac{q_i}{D}\right); \medskip \\
a_{i-1}^{i} = \disp a_{i+1}^{i} = -2\left(\frac{2}{h^2} + S_n\right); \medskip \\
a_{i-2}^{i} = \disp a_{i+2}^{i} = \frac{1}{h^2}
\end{array}
\ee
The explicit form of last row of $A$ is
\be
\begin{array}{l}
a_{N-4}^{N} = \disp \frac{3}{2h}; \medskip \\
a_{N-3}^{N} = \disp \frac{-14}{2h};\medskip \\
a_{N-2}^{N} = \disp \frac{1}{2h}\left(24 - h^2 S_n(2 - \mu)\right); \medskip \\
a_{N-1}^{N} = \disp \frac{1}{2h}\left(-18 + 4h^2 S_n(2 - \mu)\right); \medskip \\
a_{N}^{N} = \disp \frac{1}{2h}\left(5 - 3h^2 S_n(2 - \mu)\right)
\end{array}
\ee

\end{appendix}

%\vspace{4cm}

\end{document}